\documentclass[pre,superscriptaddress,showpacs,nofootinbib,
  floatfix,preprint]{revtex4-1}

\usepackage[utf8]{inputenc}
\usepackage{amsfonts}
\usepackage{amsmath}
\usepackage{amssymb}
\usepackage{tikz}
\usepackage{graphicx}
\usepackage{color}
\usepackage{epstopdf}
\usepackage{hyperref}
\usepackage{bm}

\newcommand{\cch}[1]{\left[#1\right]}

\newcommand{\prt}[1]{\left(#1\right)}
\newcommand{\aver}[1]{\left\langle #1 \right\rangle}

\hypersetup{colorlinks=true, citecolor=blue,
  pagecolor=blue,
  urlcolor=black,
  pdfcreator={pdflatex},
}

\begin{document}

\title{Influence of disordered porous media in the anomalous properties of
  a simple water  model}

\author{A. P. Furlan \footnote[1]{email - furlan@if.ufrgs.br}}
\affiliation{Instituto de Física, Univeridade Federal do Rio Grande do Sul,
  Caixa Postal 15051, 91501-570, Porto Alegre, RS, Brazil.}

\author{Carlos E. Fiore \footnote[2]{email - fiore@if.usp.br}}
\affiliation{Instituto de Física, Universidade de São Paulo,
  Caixa Postal 19044, 81531 São Paulo, SP, Brazil.}

\author{M. C. B. Barbosa \footnote[3]{marcia.barbosa@ufrgs.br}}
\affiliation{Instituto de F\'isica, Univeridade Federal do Rio Grande do Sul,
  Caixa Postal 15051, 91501-570, Porto Alegre, RS, Brazil.}

\date{\today}

\begin{abstract}
The thermodynamic, dynamic and structural behavior of a water-like system 
confined in a matrix is analyzed for increasing confining geometries. The 
liquid is modeled by a two dimensional  associating lattice gas model that 
exhibits density and diffusion anomalies,
in similarity to  the anomalies present in liquid 
water. The matrix is a triangular lattice in which fixed obstacles impose 
restrictions to the occupation of the particles. We show that obstacules
shortens all lines, including the phase coexistence, the
critical and the anomalous lines. The inclusion
of a very dense matrix not only suppress the anomalies
 but also the liquid-liquid critical point.
\end{abstract}

\maketitle
\section{Introduction}\label{sec:introduction}
The phase behavior of systems as particles interacting via the so-called
core-softened (CS) potentials has received a lot of attention recently. They 
show a repulsive core with a softening region when particles are very close
and an attractive region when particles are more distant. These CS can be 
modeled as continuous potentials or lattice gas models. For the lattice 
structure the two competing scales arise from two equilibrium configurations: 
low density and high density. This procedure generates models that are 
analytically and computationally tractable and that one hopes are capable of 
retaining the qualitative features of the real complex systems. The physical 
motivation behind these studies is the assumption that two length scales 
systems exhibit the same anomalous behaviors present in water.

Confirming this hypothesis a number of continuous~\cite{He70,De91,jagla,Ca03,
netz,Fo08} and lattice gas models~\cite{Bell:JPA3:70,Bell:JPC6:73,Fr02,Bu03,
vera,vera2,noe,Aline:PCM19:07,SZOR:130:09} show the presence of density, 
diffusion and structural anomalous behavior as observed  in 
water~\cite{kell:1975,angel:3063}.

In addition to the seventy-two anomalies~\cite{chaplin}, water has at very 
low temperatures two coexisting amorphous phases with distinct densities: the 
low density amorphous  (LDA) and high density amorphous phases (HDA) 
\cite{PhysRevLett.85.334,mishima2,PhysRevE.48.3799,mishima:1994,Nature.360.324}.
  These two amorphous phases led to the hypothesis of the existence at higher 
temperatures of two liquid phases: a low density liquid and high density liquid 
phases. Such conjecture establishes that the coexistence between these two 
liquid phases   ends in a  second critical point or also called, liquid-liquid 
critical point (LLCP)~\cite{Nature.360.324}. Experiments for testing the 
existence of this criticality are difficult since the region in the pressure 
versus temperature phase diagram where the alleged critical point exists is 
locate beyond the homogeneous nucleation limit. In order to circumvent this 
difficulty for testing the existence of the liquid-liquid critical point 
recently confined geometries have been employed~\cite{Xu05,Ch06}. In these
nanoconfined geometries the disruption of the hydrogen bonds suppress the
solidification of the system and allows for maintaining the system liquid in 
temperatures in which otherwise would be solid~\cite{Xu05,Ch06}. These 
experimental systems show convincing evidences that water exhibits two liquid 
structures at low temperatures.

The use of confining water, however brings another set of issues. What 
guarantees that the same thermodynamic and dynamic anomalies and criticality 
present in the bulk are not destroyed as the system is confined? Can confinement 
bring up new phenomena not present in the bulk system? In order to answer these 
questions water-like   atomistic or continuous effective potential models were 
explored confined by different geometries such as plates
\cite{bordin,leandro,leandro2,kumar,scheidler,meyer,giovambattista}, one pore
\cite{bordin2,bordin3,bordin4,bordin5,corradini,gallo1,gallo2,gallo3,gallo4} and
disordered matrix \cite{gallo1,strekalova,strekalova1,bonnaud,pizio,page}.
These simulations show that confinement leads to a controversial result 
regarding the melting line. While results for SPC/E water show that the melting 
temperature for hydrophobic plates is lower than the bulk and higher than for 
hydrophilic walls, for mW model no difference between  the melting temperature 
due to the hydrophobicity~\cite{Mo12} is found.

In confined systems  the TMD occurs at lower temperatures for hydrophobic 
confinement~\cite{Gi09, Ku05} and at higher temperatures for hydrophilic 
confinement~\cite{Ca09} when compared with the bulk. The diffusion coefficient, 
$D$, in the direction parallel to the plates exhibit an anomalous behavior as 
observed in bulk water. However the temperatures of the maximum and minimum of 
$D$ are lower than in bulk water~\cite{Ku05}. In the direction perpendicular
to the plates, no diffusion anomalous behavior is observed~\cite{Ha08}.

In addition to the usual density and diffusion anomalous behavior, these 
confined systems show a  variety of new effects not present in the bulk. For 
example, fluids confined in carbon nanotube exhibit formation of layers, 
crystallization of the contact layer \cite{cui,jabbarzadeh} and a superflow not 
present in macroscopic confinement \cite{chen,qin}.

The confinement by a pore, within plates or nanotubes is symmetric and even 
though it introduces a breaking of the water hydrogen bond network, this is 
done in an ordered way. Confining matrix such as the ones present in plants 
and underground water are not ordered. Recently the effects of an water-like 
liquid confined in a disordered matrix have been analyzed using a model in which
 the bonds are introduced by the inclusion of Potts variables~\cite{strekalova1}.
This study shows that the liquid-liquid coexistence line is affected by the 
increase of the density of random porous in the matrix without disappearing.

In all these studies, however the liquid-liquid transition is preserved and the 
density of confining matrix is not very high. Here, we give a further step by 
investigating effects imposed by disordered porous when the random matrix 
exhibits a high density.  The system is defined in a triangular lattice where 
the obstacles are fixed and randomly distributed. The fluid is modeled as a 
Associating Lattice Gas Model~\cite{vera} defined an occupational variable
together with a bond orientational variable. This model in the bulk shows the 
density and diffusion anomalies present in water and the liquid-gas and 
liquid-liquid criticality~\cite{vera}. Here we explore the effect in the 
chemical potential versus temperature phase diagram of the presence of the 
random fixed obstacles. 

This paper is organized as following. In Sec. \ref{sec:the_model} we present the 
model used here. In Sec. \ref{sec:sim_details} outline details of Monte Carlo 
simulations and how was calculated the thermodynamic properties of system.  
Results are presented in Sec. \ref{sec:results}, followed by conclusions in
Sec. \ref{sec:conclusions}.

\section{The Model}\label{sec:the_model}
\begin{figure}[!htb]
  \centering
  \includegraphics[scale=0.8]{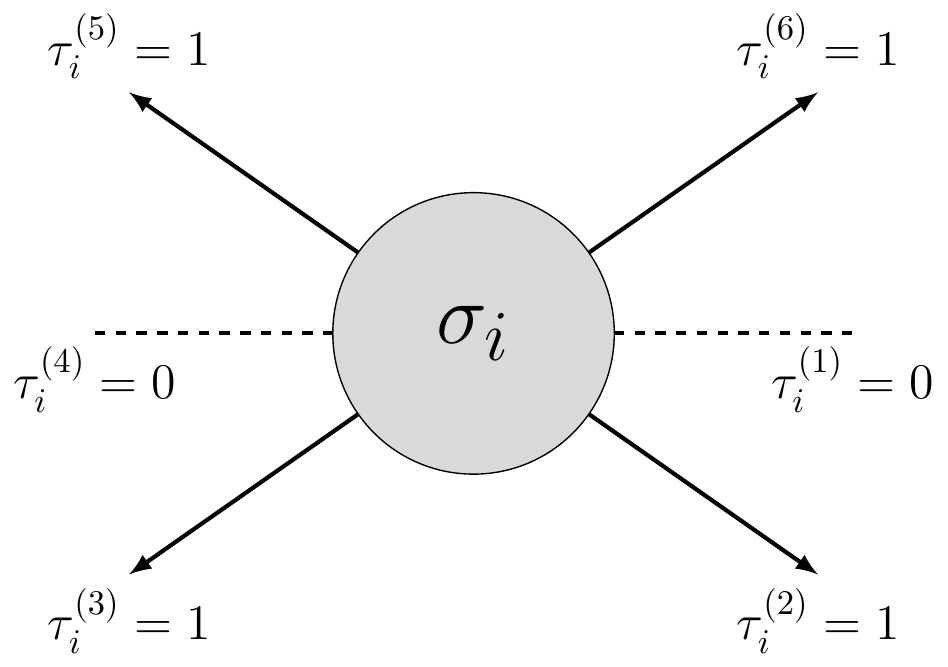}
  \caption{The occupational and orientational states of a water molecule
placed at the site $i$. In such example, the arms variables of molecule 
read: $\tau_i^{(1)}=0$, $\tau_i^{(2)}=1$, $\tau_i^{(3)}=1$,  
$\tau_i^{(4)}=0$, $\tau_i^{(5)}=1$, $\tau_i^{(6)}=1$.}
  \label{fig:state}
\end{figure}

\begin{figure}[!htb]
  \includegraphics[scale=1.2]{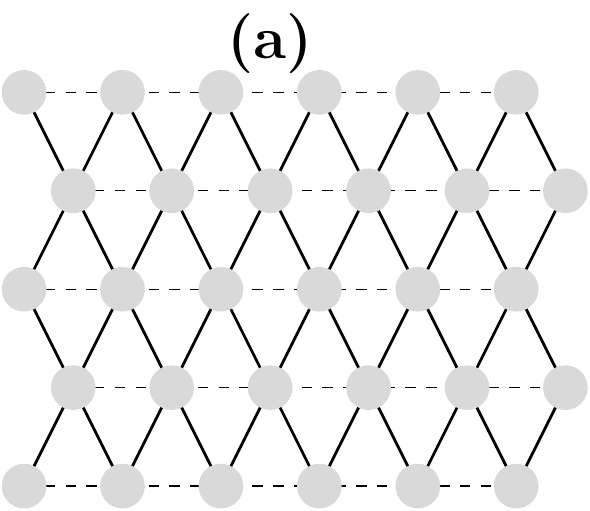} \hspace{0.6cm} \includegraphics[scale=1.2]{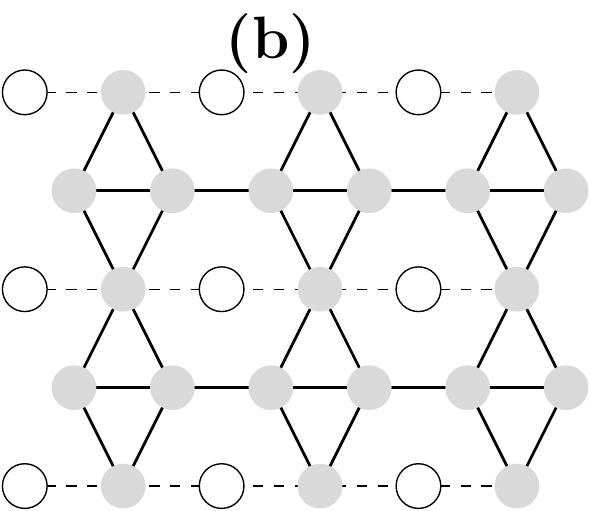}
    \caption{(a) HDL phase for the bulk system. The solid lines
indicate the bonding arms (b) LDL phase for the bulk system. The solid 
lines indicate the bonding arms. }
\label{fig:phases}
\end{figure}

\begin{figure}[!htb]
\includegraphics[scale=0.8]{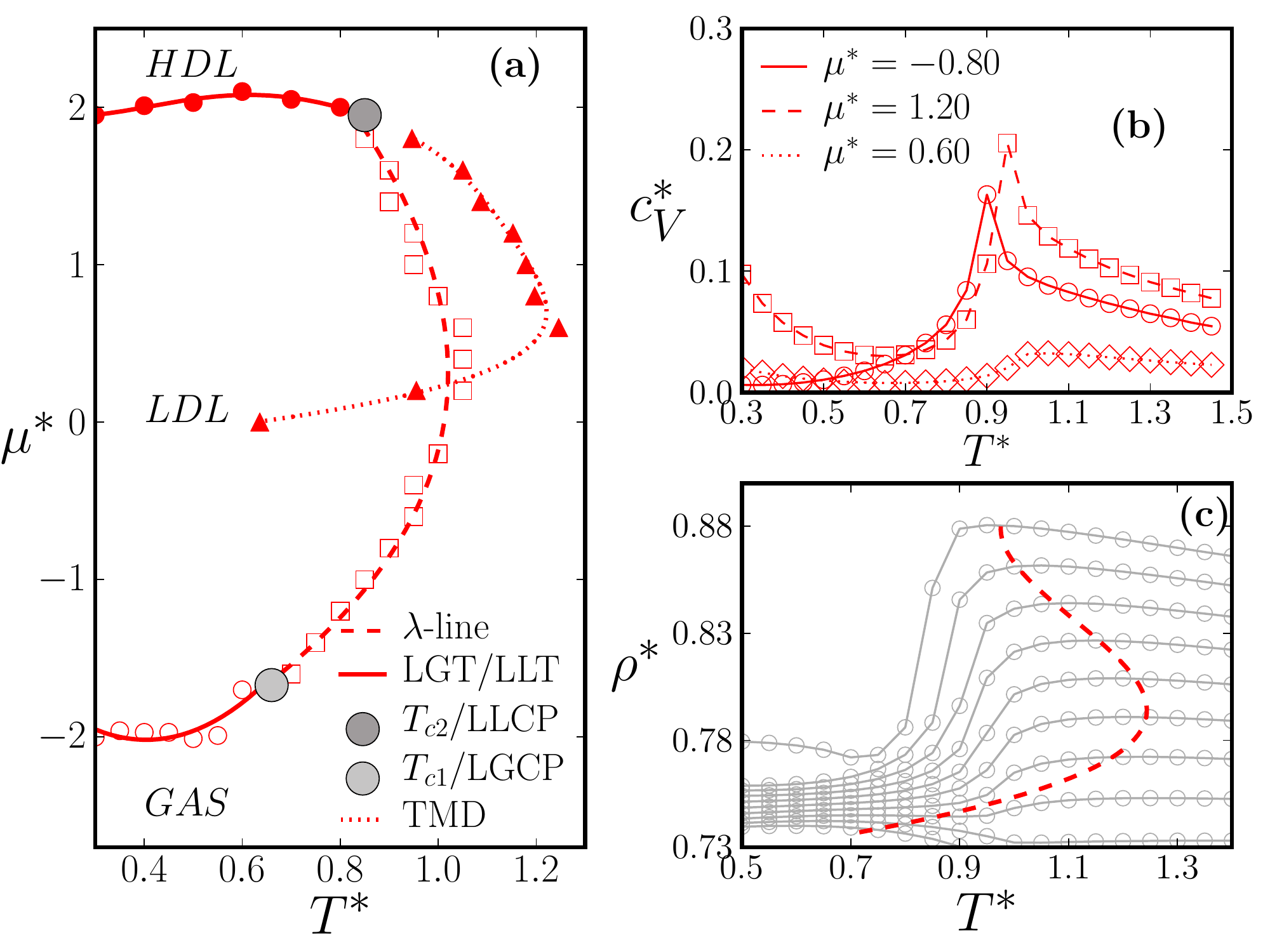}
\caption{ For the bulk System, panel (a) shows 
the phase diagram  $\mu^*$ vs $T^*$, illustrating the gas-LDL
(empty circles) and the LDL-HDL (filled circles) phase transitions, 
the $\lambda$-line (empty squares) and the TMD line (filled triangles). 
In panel (b) we plot the $c_{V}^*$ versus $T^*$ for 
$\mu^*=-0.80$ (circles), $\mu^*=0.60$ (diamonds) and
$\mu^*=1.20$ (squares). In (c) the system density $\rho$ versus 
$T^*$ for fixed $\mu^*$along the  TMD line (dashed line).}
\label{fig:muxt_cv_0}
\end{figure}

\begin{figure}[!htb]
  \centering
  \includegraphics[scale=0.9]{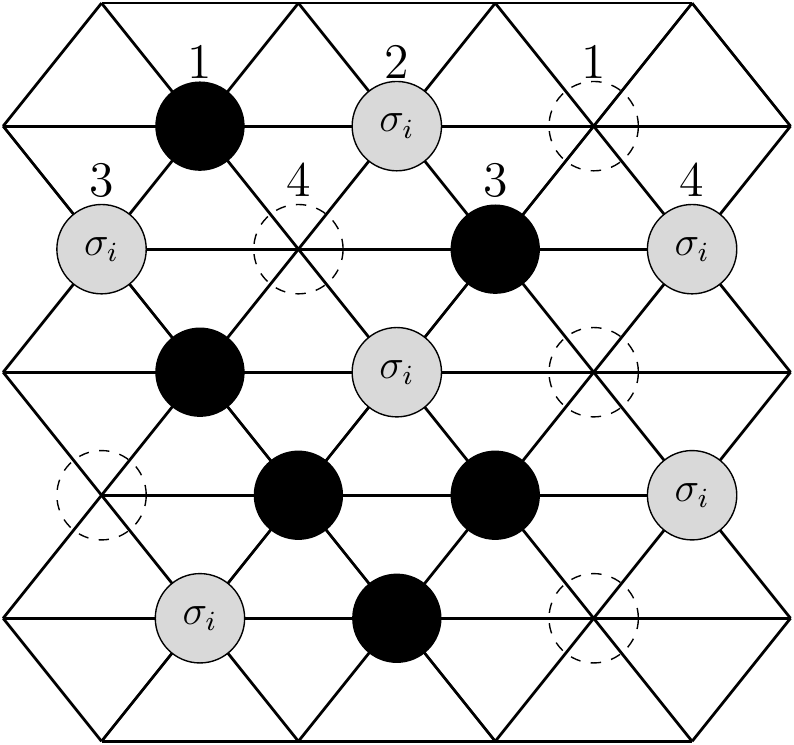}
  \caption{Example of a lattice configuration. Gray, black  and dashed circles
    describes water molecules, obstacles and empty sites, respectively. For clarity,
    hydrogen bonds have not explicitly shown.}
  \label{fig:lattice}
\end{figure}

 The model system is defined in a triangular lattice. Each accessible site $i$ 
can be empty or occupied by a water molecule. Empty sites have $\sigma_i =0$ 
while occupied  sites have $\sigma_{i} =1$. Each water molecule has orientational
states represented by the variable $\tau$ that presents six arms, being two 
inert arms with $\tau_i=0$ and  four active arms with $\tau_{i}=1$. They 
represent the possibility of a molecule to form  hydrogen bonds with up to four 
neighbor molecules. For representing the symmetry present in water,  two inert 
arms are diametrically positioned and just three different orientational states 
are possible. Fig. \ref{fig:state} exemplifies the geometry of a water molecule.
A hydrogen bond is formed only when  the active arms of two neighbor  molecules 
point out to each other, $\tau_i\tau_j =1$. In this case,  the interaction energy between two bonded arms reads $-v$ and while non bonded arms contribute with a 
higher energy of $-v+2u$ (punishement for non forming hydrogen bonds). The 
Hamiltonian of the system is given by
\begin{equation}
  {\cal H} = 2u\sum_{\aver{i,j}} \sigma_{i}\sigma_{j}\cch{
    \prt{1-\frac{v}{2u}}-\tau_{i}\tau_{j}} - \mu\sum_{i}\sigma_i.
\end{equation}
The phase behavior of the system in the absence of obstacles was already 
analyzed~\cite{Aline:PCM19:07}. Below it will be reviewed. At ground state, 
$T^*\equiv T/v=0$, the grand potential per site is $\Phi=e-\mu N$ where 
$e=\aver{\cal H}/L^2$. For low chemical potentials, the lattice is empty and 
the system is constrained in gas phase, $\rho=0$. In this phase the grand 
potential is $\Phi_{GAS}=0$. Increasing the chemical potential the system 
reaches a point at which the gas phase coexists with a low density liquid phase 
(LDL). In this phase, the density is $\rho=3/4$ and each particle forms four 
hydrogen bonds with its neighbor, resulting in a grand potential per site 
$\Phi_{LDL}/L^2=-(3/2)v-(3/4) \mu$ and consequently in a gas-LDL coexistence 
chemical potential $\mu_{G-LDL}^*=\mu_{G-LDL}/v=-2$. For high chemical potentials, 
all sites of lattice are occupied by particles, resulting in a density $\rho=1$ 
and  grand potential per site $\Phi_{HDL}/L^2=-3v+2u-\mu$. The coexistence
between the LDL phase and the HDL phase occurs at 
$\mu_{LDL-HDL}^*=\mu_{LDL-HDL}/v=8u/v-6$. The main features of LDL and HDL phases 
are exemplified in Fig. \ref{fig:phases} for two possible configurations at 
$T^*=0$.
At temperatures $T^*>0$, the model was studied by Monte Carlo simulations
whose  phase diagram is shown in Fig.
\ref{fig:muxt_cv_0}.
The gas-LDL and LDL-HDL transition lines are first-order transitions, ending in
respective  tricritical points $T_{c1}$ and $T_{c2}$, respectively that are joined 
by a line of continuous transitions, called  $\lambda-$line. For the bulk case,
$T_{c1}$ and $T_{c2}$ read $0.65$  and $0.825$, respectively. In order to 
understand the differences between the LDL and HDL phases, we divide the lattice 
in four sublattices as illustrated in the Fig.~\ref{fig:lattice}. The LDL phase 
is characterized by one of the sublattices being empty while all the others are 
filled, in such a way that  the transition to the HDL phase occurs when the 
empty sublattice is filled. Also, it is  signed by a rotation in the inert arms, 
in which  in the HDL phase they are all parallel. In addition, in the LDL phase 
each particle  forms four bonded arms that show a zigzag structure, whereas in  
the HDL phase each particle also forms four bonded arms but in addition to the 
zigzag structure two parallel lines appear. In the regime of very high 
temperatures, the system is disordered, in which the sublattice occupations do 
not exhibit any ordering. By lowering $T$, the $\lambda$-line is crossed, which 
one sublattice is emptied and the others  remaining filled with an 
reorganization of the inert arms that form the above ordered zig-zag structure.

The density of bonds, $\rho_{hb}=
\dfrac{1}{L^2}\sum_{i=1}^{L^2} \sum_{i+\delta}\sigma_i\sigma_{i+\delta}\tau_{i}\tau_{i+\delta}$
is also an important quantity for characterizing the phase transitions. At 
$T^*=0$  the gas, LDL and HDL phases has $\rho_{hb}$ reading $0,1.5$ and $2$, 
respectively. Thus the phase transitions are also signed by changes in the 
fraction of hydrogen bonds. At high temperature the system is disordered and 
the $\lambda-$line  is obtained through the specific heat at constant volume 
$c_V$ by
\begin{equation}
  \label{eq:spec_heat}
  c_{V}=\dfrac{1}{VT^{2}}\cch{\aver{\delta{\cal H}^2}_{\mbox{\tiny gcan}} -
  \dfrac{\aver{\delta {\cal H}\delta N}^2_{\mbox{\tiny gcan}}}{\aver{\delta N^2}_
  {\mbox{\tiny gcan}}}} + \dfrac{3Nk_B}{2V}
\end{equation}
where  $\delta X=X-\aver{X}$ with $X={\cal H}$ and $N$ and averages are
evaluated in the ensemble of $T,\mu$ fixed.

In this work the porous matrix is introduced by considering fixed obstacles 
that are randomly distributed in the lattice. Each obstacle occupies a single 
site and interacts with the particles  via a ``hard core'' constraint. 
The density of obstacles is given by $\rho_o=N_o/L^2$ where $N_o$ is the number 
of obstacles and $L^2$ is the system volume. In Fig. \ref{fig:lattice},
we exemplify a lattice configuration composed of water, obstacles  and empty 
sites.

\section{The Methods and Simulation Details}\label{sec:sim_details}
All the thermodynamic  properties have been obtained by performing grand 
canonical Monte Carlo (MC)  simulations for fixed $T^*,\mu^*$ and $\rho_0$ 
\cite{Frenkel:book:2002}. Microscopic configurations are generated according 
to the Metropolis algorithm~\cite{Metropolis:JCP:1953} described as follows.
Obstacles are initially randomly distributed and a given site $k$ other than 
a porous is randomly chosen. Two sorts of transitions are possible, taking 
into account $k$ be empty or occupied by a water molecule. In the former case, 
a water molecule in one of its arm states is chosen, whereas in the latter
  one of the  three possibilities (including the other two arm states  and 
empty site) are performed. All of possible transitions are chosen with equal 
probability. Next, we evaluate the energy difference $\Delta {\cal H}$ between 
the original and the new configuration. The configuration change is accepted 
according to the Metropolis prescription ${\rm min}\{1, e^{-\beta \Delta {\cal H}} \}$,
where $\beta=1/k_{B}T$. A Monte Carlo step is defined as the number of trials 
for generating new configurations. After discarding a sufficient number of 
initial MC steps, relevant quantities are evaluated. In addition to the 
thermodynamic quantities, we also investigated the influence of obstacles in 
the dynamic properties, characterized through the diffusion coefficient $D$ 
given by Einstein's relation
\begin{equation}
D=\lim_{t \rightarrow \infty} \frac{\aver{\Delta r(t)^2}}{4t},
\label{eq1}
\end{equation}
where $\langle \Delta r(t)^{2}\rangle=\langle (r(t)-r(0))^{2} \rangle$ is the 
mean square displacement per particle and time is measured in Monte Carlo steps. 
The numerical MC procedure for calculating the diffusion is described as follows.
First, the system is equilibrated by employing the previous Metropolis dynamics 
for fixed $T^*$ and $\mu^*$. After the equilibrium is reached,  an occupied site 
$i$ and it's neighbor $j$ are chosen randomly. In case of neighbor site $j$ be 
empty, the molecule moves to the empty site also following  the above Metropolis 
prescription ${\rm min}\{1,e^{-\beta \Delta {\cal H}} \}$, where $\Delta {\cal H}$ is 
the difference of energy due to the movement. A Monte Carlo step is defined 
through the number of trials of movement for every particle. After repeating this
 algorithm $nt$ times, where $n$ is the number of molecules in the lattice, the 
diffusion coefficient is calculated from Eq.(\ref{eq1}).

Numerical simulations have been performed for triangular lattices of size $L=35$ 
and periodic boundary conditions for three representative values of density of 
obstacles $\rho_o=0.08,0.24$ and $0.40$ have been considered. In all cases, we 
have used $10^6$ Monte Carlo (MC) steps to equilibrate the system and $10^6$ MC 
steps for evaluating the relevant quantities.

\section{Results}\label{sec:results}
\subsection{Structural and thermodynamic behavior}

\begin{figure}[!htb]
\centering
\includegraphics[scale=0.8]{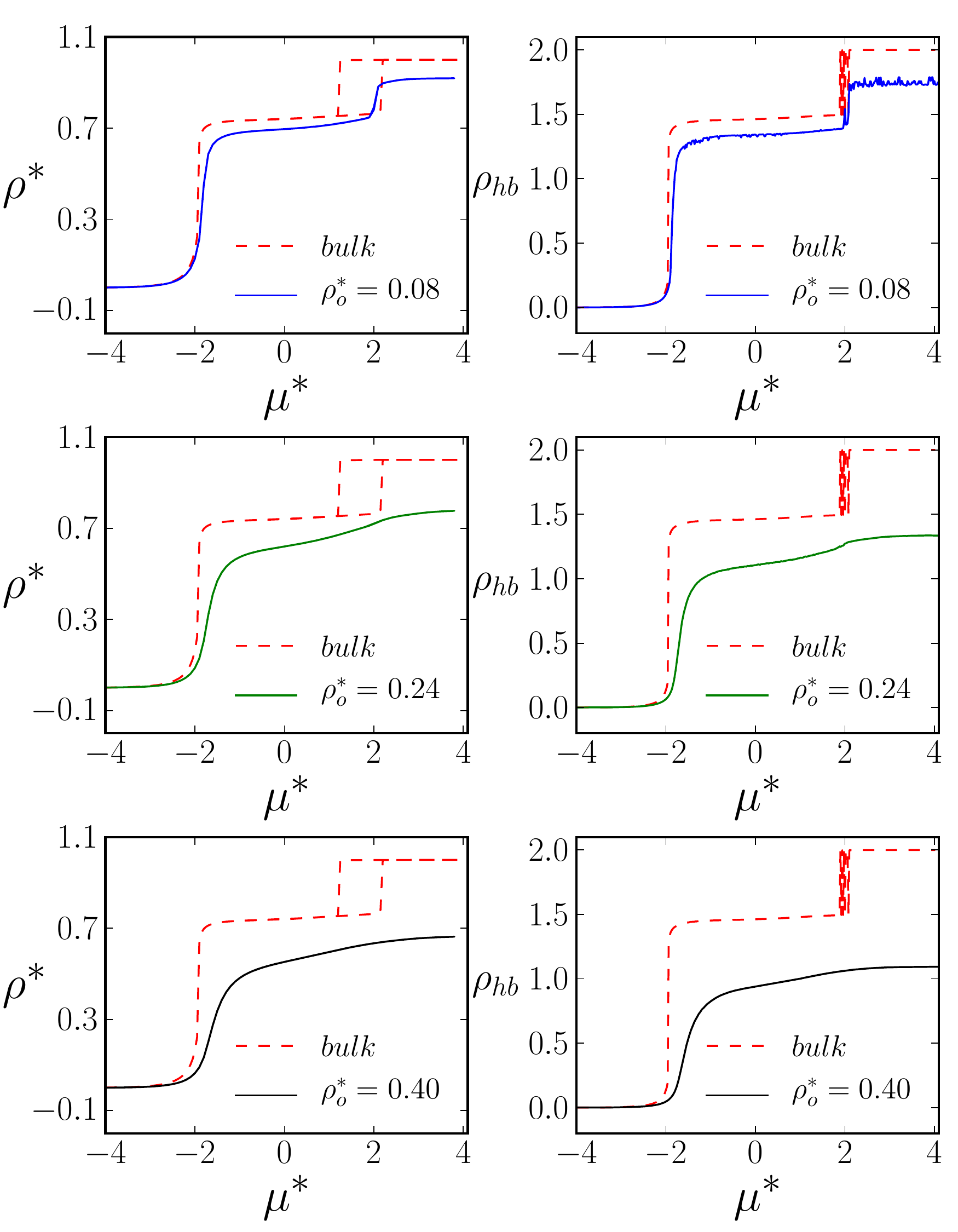}
\caption{$\rho$ vs $\mu^*$ for distinct porous densities $\rho_o$ for $T^*=0.40$.}
\label{fig:fig7}
\end{figure}

\begin{figure}[!htb]
  \centering
  \includegraphics[scale=0.8]{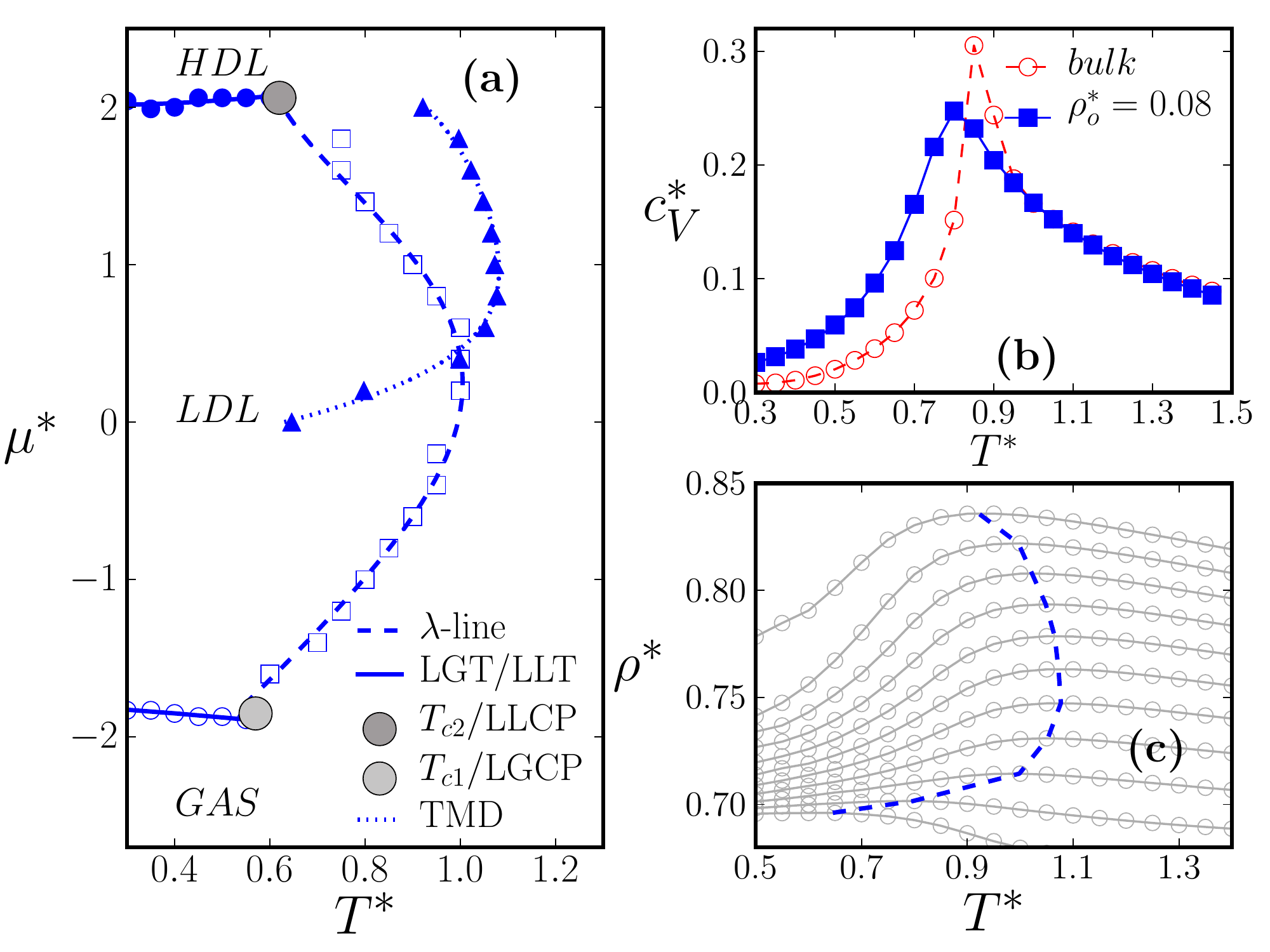}
  \caption{For $\rho_o=0.08$, in  $(a)$ the phase diagram in the plane of
reduced chemical potential $\mu^*$ versus reduced
temperature $T^*$. Empty and filled circles denote the  Gas-LDL  and the 
LDL-HDL  phase transitions, respectively.  The $\lambda$ and TMD lines are 
described by empty squares filled triangles, respectively. Panel $(b)$ shows 
the specific heat at constant volume $c_V$ versus $T^*$ for the system with 
obstacles (filled squares) and bulk system (empty circles) at $\mu =-1.00$.   
Panel $(c)$ shows $\rho$ versus $T^*$ for $\mu^*=0.0,\ldots,2.2$ showing the TMD 
line (dashed line).}
\label{fig:muxt_cv_100}
\end{figure}

\begin{figure}[!htb]
  \centering
  \includegraphics[scale=0.7]{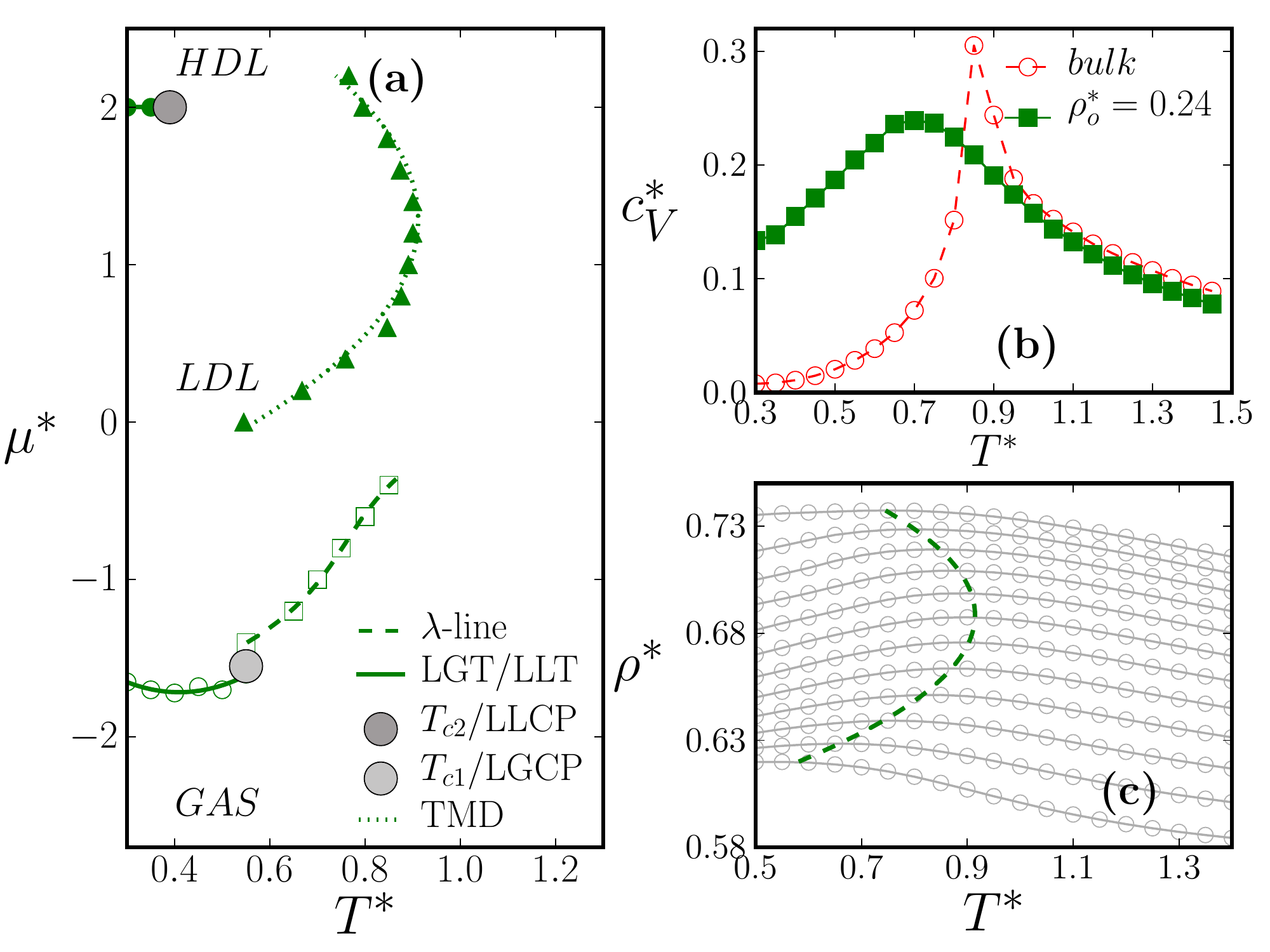}
  \caption{For $\rho_o=0.24$, in  $(a)$ the phase diagram in the plane of
reduced chemical potential $\mu^*$ versus reduced temperature $T^*$. Empty 
and filled circles denote the  Gas-LDL  and the LDL-HDL  phase transitions,
respectively.  The $\lambda$ and TMD lines are described by empty squares 
filled triangles, respectively. Panel $(b)$ shows the specific heat at 
constant volume $c_V$ versus $T^*$ for the system with obstacles (filled squares) 
and bulk system (empty circles) at $\mu =-1.00$.   Panel $(c)$ shows $\rho$ 
versus $T^*$ for $\mu^*=0.0,\ldots,2.0$ showing the TMD line (dashed line).}
\label{fig:muxt_cv_300}
\end{figure}

\begin{figure}[!htb]
  \centering
  \includegraphics[scale=0.8]{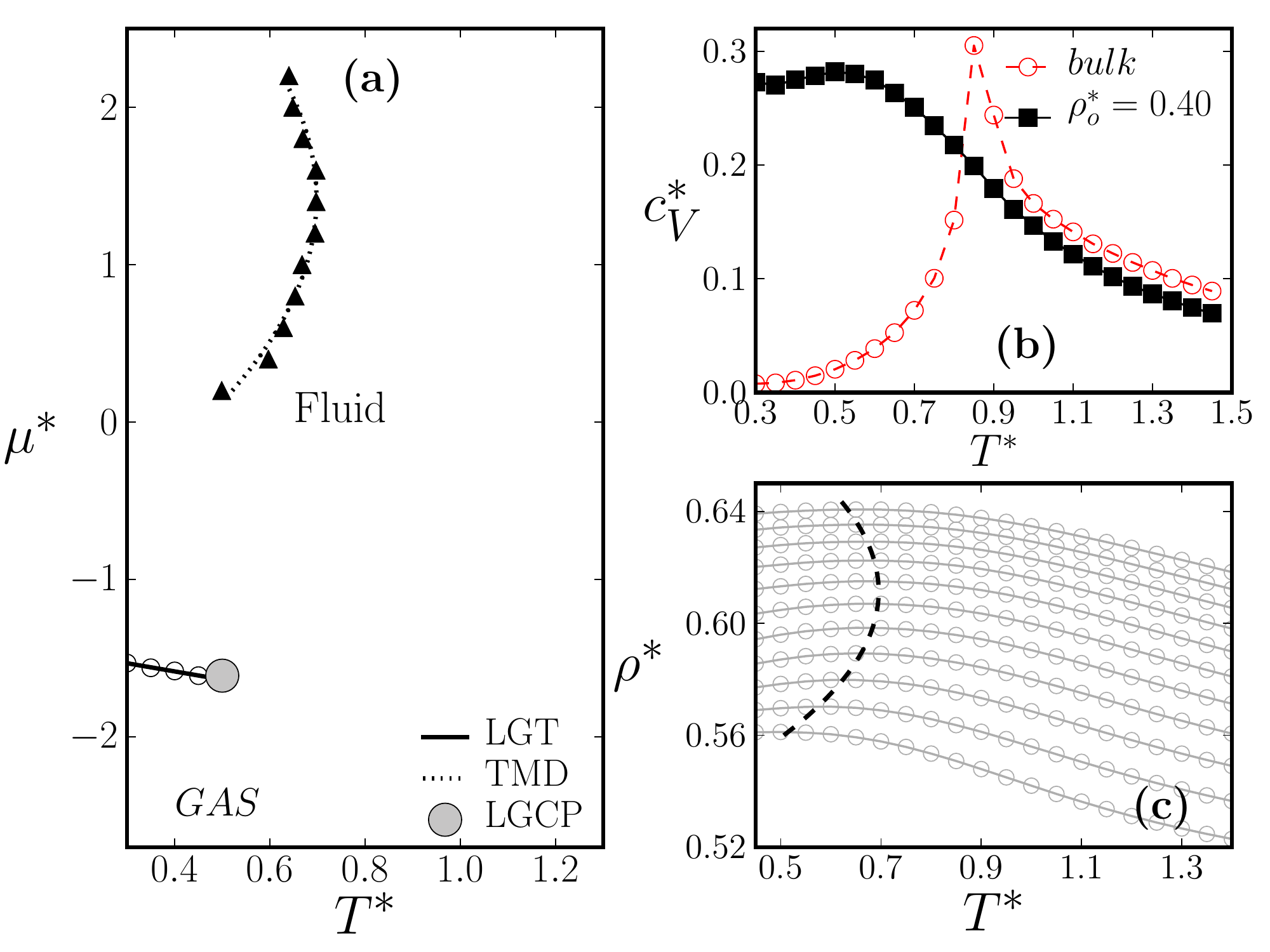}
  \caption{For $\rho_o=0.40$, in  $(a)$ the phase diagram in the plane 
of reduced chemical potential $\mu^*$ versus reduced temperature $T^*$. 
Empty and filled circles denote the  Gas-LDL  and the LDL-HDL  phase 
transitions, respectively. The $\lambda$ and TMD lines are described by 
empty squares filled triangles, respectively. Panel $(b)$ shows the specific 
heat at constant volume $c_V$ versus $T^*$ for the system with obstacles 
(filled squares) and bulk system (empty circles) at $\mu =-1.00$. Panel $(c)$ 
shows $\rho$ versus $T^*$ for $\mu^*=0.2,\ldots,2.2$ showing the TMD line 
(dashed line).}
\label{fig:muxt_cv_500}
\end{figure}

First, let us exam what happens with the phases present in the bulk system as 
the obstacles are introduced. Fig. \ref{fig:fig7} shows the water density 
$\rho$, versus the reduced chemical potential  $\mu^*$, for distinct porous
densities at the fixed temperature $T^*=0.40$. The inclusion of obstacles changes
 the gas-LDL and the LDL-HDL phase transition, whose effect becomes more 
pronounced as $\rho_o$ increases. In particular, by raising the porosity, the 
density gap between the liquid phases becomes less abrupt and inclusion of 
obstacles move the transition points for larger chemical potentials.
Figs.~\ref{fig:muxt_cv_100}, \ref{fig:muxt_cv_300} and \ref{fig:muxt_cv_500}
illustrate the chemical potential versus temperature phase diagrams for 
$\rho_o=0.08, 0.24$ and $0.40$, respectively. In particular, by increasing  
$\rho_o$ the tricritical points $T_{c1}$ and $T_{c2}$, in which the gas-LDL
and LDL-HDL  coexistence lines meet the $\lambda-$line, decreases as shown in 
Figs.~\ref{fig:muxt_cv_100} , \ref{fig:muxt_cv_300} and \ref{fig:muxt_cv_500}.
More specifically, while the bulk gas-LDL tricritical point is located at 
$T_{c1}=0.65$, it moves to $T_{c1}=0.60, 0.55$ and $0.52$ for $\rho_o=0.08,0.24$ 
and $0.40$, respectively.

This scenario becomes even more drastic in the case of the LDL-HDL phase 
transition. The tricritical point not only decreases its value from  
$T_{c2}=0.825$ (bulk) to $T_{c2}=0.57$ and $T_{c2}=0.52$ for $\rho_o=0.08$ and 
$0.24$, respectively but  it disappears for $\rho_o=0.40$, implying the absence
of liquid-liquid transition line. In addition the $\lambda$-line can  also be 
found only for $\rho_o=0.08, 0.24$, in which for $\rho_0=0.40$ the original bulk 
tricritical $T_{C1}$ becomes a single critical point
.

\begin{figure}[!htb]
    \includegraphics[scale=0.35]{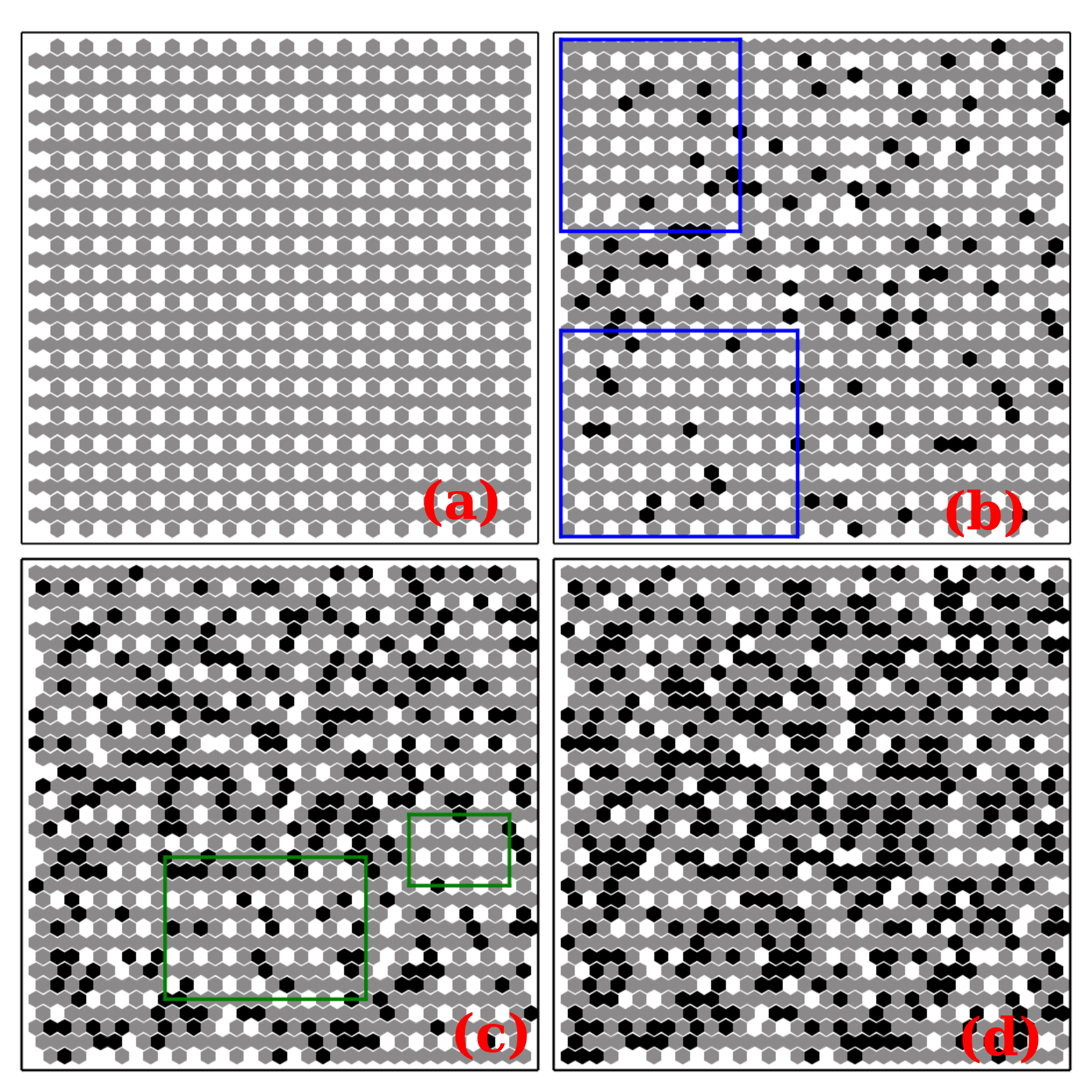}
    \caption{Spatial snapshot (35 $\times$ 35 sites) of triangular
      lattice. Each site is represented by hexagon, with its six
      nearest-neighbor sites. White hexagons represent vacancies,
      black represent obstacles and gray represent water-like particles.
      The snapshots exhibit character configurations of system with
      chemical potential $\mu^*=-0.5$ and temperature $T^*=0.3$. In (a)
      we present the bulk system. In (b) the system submitted at
      low degree of confinement $\rho_o=0.08$ and the blue rectangles
      denote the regions where the characteristic geometry of LDL of ALG
      is preserved. In (c), intermediate degree of confinement $\rho=0.24$,
      and green rectangles denote the LDL structure. The highest degree
      of confinement $\rho=0.40$ is shown in (d).}
    \label{fig:snaps}
\end{figure}

The  above changes in liquid phases as well as transition points can be 
understood by verifying that the inclusion of porous suppress partially the 
structured patterns found in the LDL and HDL phases (see e.g  Fig. 
\ref{fig:phases}(a) and (b) for the bulk case).  In the case of the LDL phase  
the ordered  bulk structure  is distorted as $\rho_0$ increases, as illustrated
 in the Fig. \ref{fig:snaps} for $\mu^*=-0.5$. For the lowest case $\rho_o=0.08$,
  the degree of confinement is low and most occupied sites preserve at least 
three bonds. As the density of obstacles is increased (here exemplified for 
$\rho_o=0.16$ and $0.24$) the fraction of disrupted bonds  increases, reaching a 
limit in which the connectivity of the  network is lost. Similar effect is 
verified in the HDL phase, but the effect is more pronounced in such case. This 
can be understood by recalling that in contrast to the HDL phase, porous occupy 
partially empty sites with neighboring molecules  not forming hydrogen bonds in 
the LDL phase. This lost of connectivity also explains, in similarity with
the decrease of tricritical points, why the transition from the disordered 
structure to the LDL through the $\lambda$-line occurs for lower temperatures 
than the bulk.  Recalling that such transition is characterized by the 
disordered phase ordering by making one of the sublattices empty,  the inclusion 
of porous makes the  system entropy be larger than the  bulk, with partial 
disruption of hydrogen bonds. Thus, all transition points,  move for lower 
temperatures as a  way for "compensating" the above increase of disorder. In 
other words, due to the inclusion of porous, the structured phases exist only 
for lower temperatures than the bulk, whose decreasing become more pronounced as
 $\rho_0$ increases. Finally, for high density of obstacles the transition is 
destroyed by  enhancement of fluctuations. The last comment concerns in the 
comparison between the TMD as $\rho_0$ increases. As for the  transition lines, 
the TMD shortens and  move for lower temperatures (with maximum $\rho$ 
decreasing) as $\rho_0$ increases. However, in contrast previous results, for 
$\rho_0=0.40$ a tiny TMD (ranged from $T^*=0.50$ to $0.70$  with $\rho=0.56$ to 
$0.64$)  is  verified.

 \begin{figure}[!htb]
\includegraphics[scale=0.80]{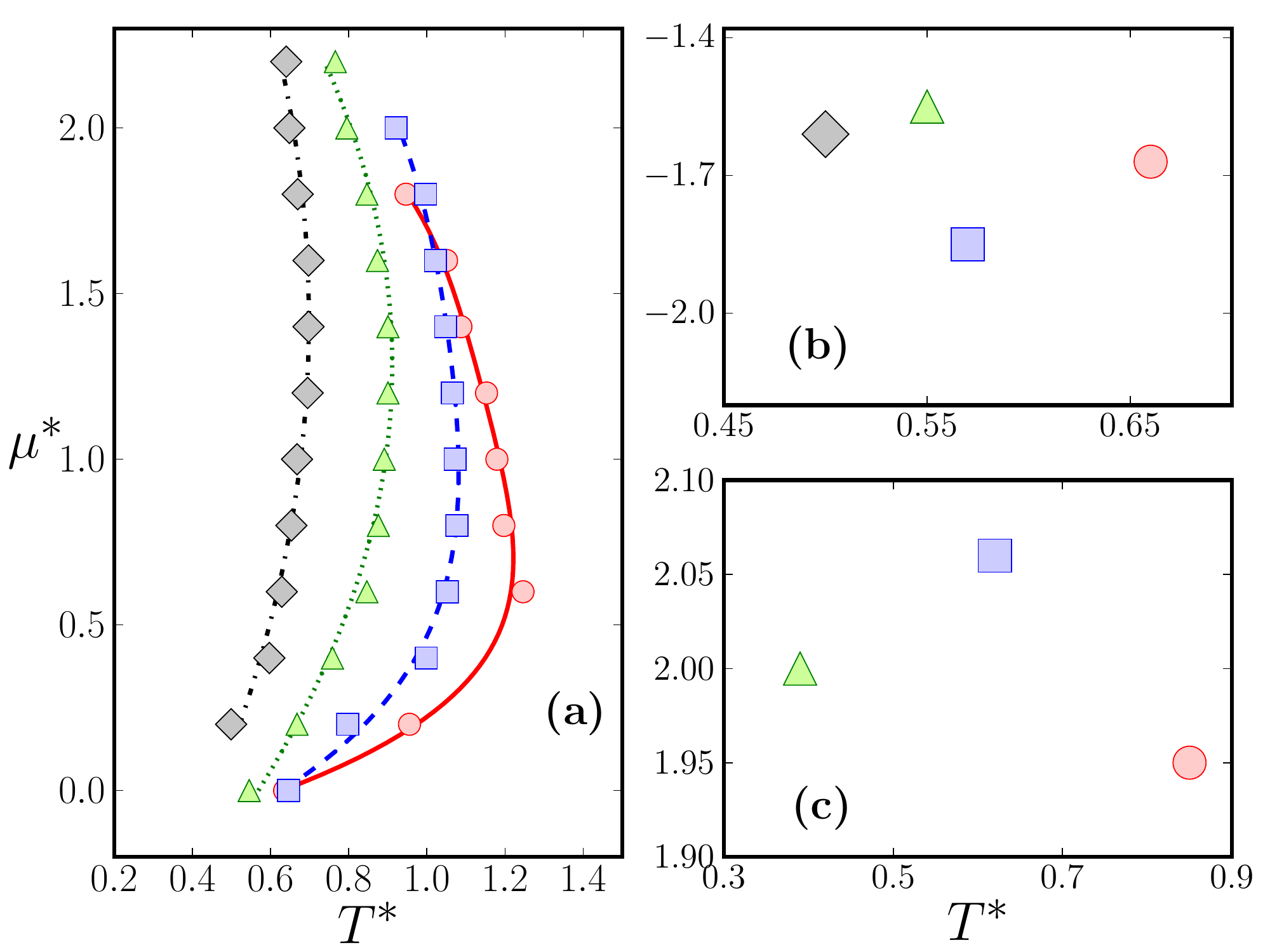}
\caption{Chemical potential versus temperature illustrating (a) the TMD lines
 (b) the gas-LDL tricritial point (c) the LDL-HDL critical point values for the bulk and
 the system with different concentrations of obstacles.}
\label{fig:tmd_cp}
\end{figure}

\subsection{Diffusion and dynamic anomaly}

 Besides the influence of immobile obstacles in the thermodynamic quantities, 
another relevant question concerns what happens with the system mobility as the
density of obstacles increases.
\begin{figure}[!htb]
\includegraphics[scale=0.80]{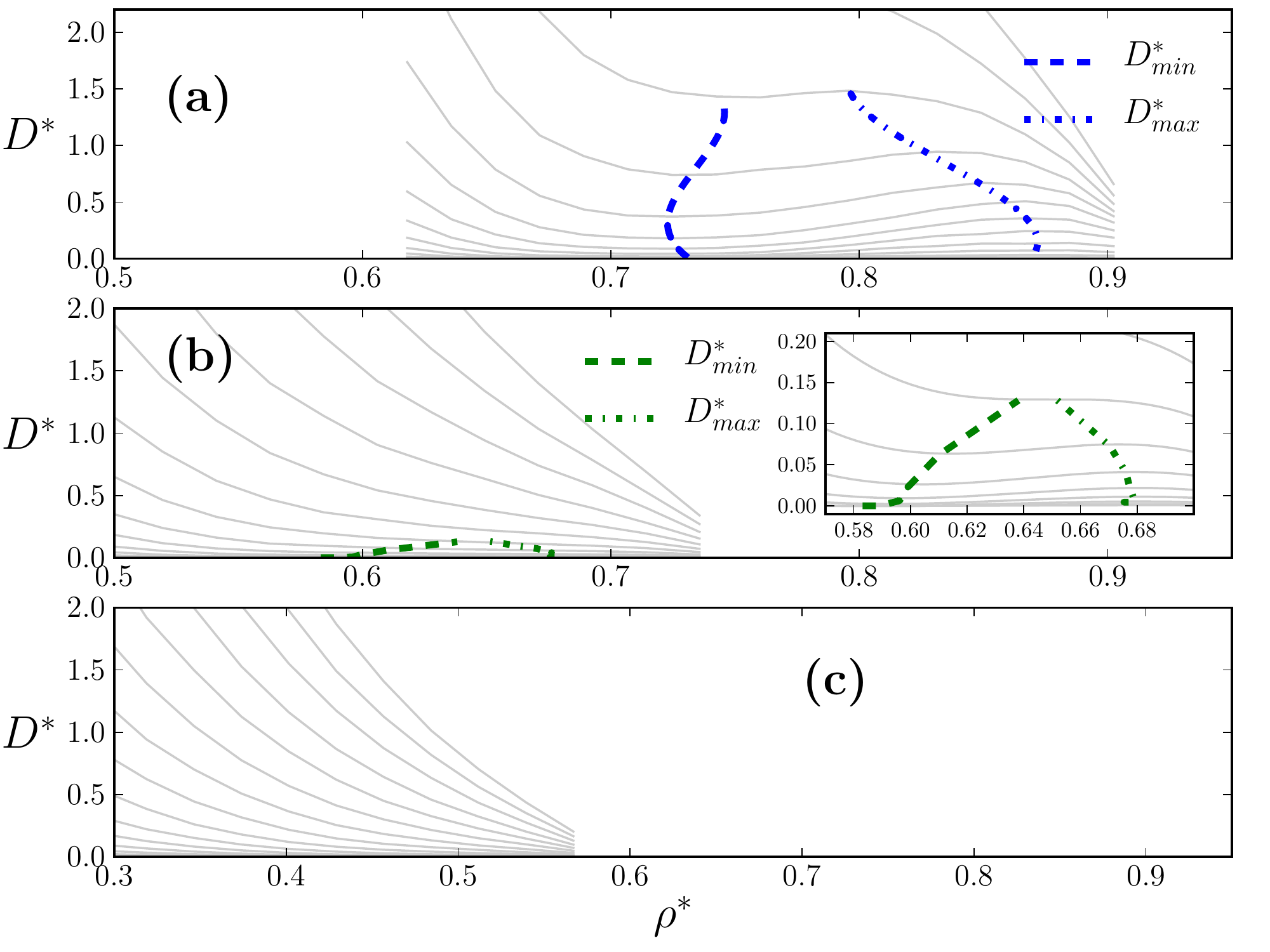}
\caption{Diffusion coefficient versus density at fixed temperature. The gray lines are
  fit of diffusion obtained by simulation. In case (A) we present the results for
  $\rho_o=0.08$, in which the gray lines are just fit of diffusion obtained by simulation
  , blue dashed and dot-dashed connect the minimum and maximum in diffusion respectively.
  In (B) we have the same plot of case (A) but for the case $\rho_o=0.24$ and on inset we
  plot a zoom of anomalous region. In (C) the results for $\rho_o=0.40$. The temperatures
  studied were $T^*=0.30 \ldots 1.00$ with $\Delta T^*=0.05$}
\label{fig:dif_r2}
\end{figure}
Fig. \ref{fig:dif_r2} shows the diffusion coefficient computed using Eq. 
(\ref{eq1}) for different $T^*$'s and $\rho_0$'s.  In similarity with the
pure model, the diffusion coefficient presents  an increasing anomalous 
behavior until a maximum value by raising $\rho$ for $\rho_0=0.08$ and  $0.24$. 
The only exception is the case with density of obstacles $0.40$.

The reason concerns that the dynamic anomaly depends crucially of the presence 
of a high  number of neighbor sites occupied by the fluid~\cite{Netz:2002}. 
The obstacles make this difficult and for a very high number of obstacles, the 
mobility becomes even impossible.

Since for  water-like systems, typically the region in the $\mu^*$-$T^*$ phase 
diagram in which the density anomaly is present is  close to the region where
the diffusion anomaly appears. Therefore one expects that the suppression of the
 first is directly related to the disappearance of the other.

\section{Conclusion}\label{sec:conclusions}

The effects of fixed obstacles in thermodynamic and dynamic properties of
an simplified water-like model have been investigated. For low degree of
confinement, the thermodynamic, structural and dynamic properties of model 
are almost totally preserved due to the low steric effects. For intermediate 
case, $\rho_o=0.24$, the system suffers significant changes such as, the decrease
of the critical and tricritical points to lower temperatures, resulting in a 
reduction of coexistence regions. This effect is more dramatic for the 
liquid-liquid coexistence that disappear for  $\rho_o=0.40$. The density and 
diffusion anomalous regions are also  shifted  to lower temperature, keeping the 
reduction in temperature-chemical potential phase diagram. The disappearance of 
the liquid-liquid  temperature also reflects in the absence of density and 
diffusion anomalous regions in the limit of large  density of obstacles. Both 
effects are related to both the entropy increase due to the presence of the 
obstacles and the disruption of the bonds network.

\section{Acknowledgments}
We acknowledge the Brazilian agency CAPES (Coordena\c c\~ao de 
Aperfei\c coamento de Pessoal de N\'{\i}vel Superior) for the financial support 
and Centro de F\'{\i}sica  Computacional - CFCIF (IF-UFRGS) for computational 
support .

\end{document}